\documentclass[12pt]{article}
\usepackage{graphicx}
\usepackage{epstopdf}
\usepackage{amsmath}
\usepackage{amsfonts}
\usepackage{amssymb}
\usepackage{enumerate}
\usepackage{amsthm}
\usepackage{array}
\usepackage{epsfig}
\usepackage[small,bf]{caption}
\usepackage{appendix}
\usepackage{dsfont}
\usepackage{hyperref}
\def\1ad{\mbox{\normalsize $^1$}}
\def\2ad{\mbox{\normalsize $^2$}}
\def\3ad{\mbox{\normalsize $^3$}}
\def\4ad{\mbox{\normalsize $^4$}}
\def\5ad{\mbox{\normalsize $^5$}}
\def\6ad{\mbox{\normalsize $^6$}}
\def\7ad{\mbox{\normalsize $^7$}}
\def\8ad{\mbox{\normalsize $^8$}}

\def\beq{\begin{equation}}                     %
\def\eeq{\end{equation}}                       %
\def\bea{\begin{eqnarray}}                     
\def\eea{\end{eqnarray}}                       
\def\dj{\hbox{d\kern-0.347em \vrule width 0.3em height 1.252ex depth
-1.21ex \kern 0.051em}}

\def\Tr{{\rm Tr\,}}

\def\ket{\rangle}
\def\bra{\langle}

\def\pt{\partial}

\def\Dirac{\,\raise.15ex\hbox{/}\mkern-13.5mu D}
\def\dirac{\,\raise.15ex\hbox{/}\kern-.57em \partial}
\def\pslash{\,\raise.15ex\hbox{/}\kern-.57em p}

\def\R{ \mathds{R}}
\begin{document}

                     %

\newcommand{\sheptitle}
{A note on the extensivity of the  holographic entanglement entropy}
\newcommand{\shepauthora}
{{\sc
 Jos\'e L.F.~Barb\'on and Carlos A. Fuertes}}

\newcommand{\shepaddressa}
{\sl
Instituto de F\'{\i}sica Te\'orica  IFTE UAM/CSIC \\
 Facultad de Ciencias C-XVI \\
C.U. Cantoblanco, E-28049 Madrid, Spain\\
{\tt jose.barbon@uam.es}, {\tt carlos.fuertes@uam.es} }

\newcommand{\shepabstract}
{ 
\noindent

We consider situations where the \emph{renormalized} geometric entropy, as defined by the AdS/CFT ansatz of Ryu and Takayanagi,
shows extensive behavior in the volume of the entangled region. In general, any holographic geometry that is `capped' in
the infrared region is a candidate for extensivity provided the growth of minimal surfaces saturates at the capping
region, and the induced metric at the `cap' is non-degenerate. Extensivity is well-known to occur for highly thermalized
states. In this note, we show that the holographic ansatz predicts the persistence of the extensivity down to vanishing
temperature, for the particular case of conformal field theories in $2+1$ dimensions with a magnetic field and/or
electric charge condensates.}

\begin{titlepage}
\begin{flushright}
{IFTE UAM-CSIC/2007-66\\
}

\end{flushright}
\vspace{0.5in}
\vspace{0.5in}
\begin{center}
{\large{\bf \sheptitle}}
\bigskip\bigskip \\ \shepauthora \\ \mbox{} \\ {\it \shepaddressa} \\
\vspace{0.2in}

{\bf Abstract} \bigskip \end{center} \setcounter{page}{0}
 \shepabstract
\vspace{2.5in}
\begin{flushleft}
\today
\end{flushleft}


\end{titlepage}

\newpage


\setcounter{equation}{0}

\section{Introduction}
\noindent

A very interesting proposal for a holographic \cite{rads} computation of geometric (entanglement) entropy \cite{enten}
in conformal field theories was put forward in \cite{ryutaka} (see \cite{lista} for subsequent work). The proposed
procedure bears a close similarity with the holographic calculation of Wilson and 't Hooft loops in gauge theories and
applies strictly to a large-$N$, strong coupling limit of the quantum field theories (QFT) in question.

In recent years, entanglement entropy has gradually emerged as a useful non-local order parameter to characterize new
phases in strongly coupled systems at zero temperature (cf. for example \cite{latorre}). It is therefore of great interest to parametrize
the different types of qualitative behavior that can be identified from the AdS/CFT side. In this direction, the
authors of \cite{kutkleb} have made the interesting observation that the scaling of geometric entanglement entropy is a
good order parameter of confinement in gauge theories that admit standard AdS-like duals (see also the earlier work of Ref. \cite{nishitaka}). In particular they found that
for regions of size bigger than the confining scale the renormalized entanglement entropy\footnote{We understand by
  renormalized entanglement entropy the quantity obtained by subtracting the ultraviolet divergent term to the
  entanglement entropy. A more refined quantity is the \emph{entropic C-function} \cite{CasiniHuerta} to which all our comments
  equally apply.} suddenly becomes strictly proportional
to the area of the region with a constant of proportionality \emph{independent} of the size, unlike conformal field
theories. In this paper, we follow this program, focusing on the property of {\it extensivity}, an admittedly {\it non
  generic} feature of geometric entropy in standard weak-coupling calculations at zero temperature.

In this note, we begin with a review of the holographic proposal of \cite{ryutaka}, emphasizing
the natural emergence of the non-extensive `area law' for the geometric entropy in models with conformal ultraviolet
fixed points. It is then argued that extensivity of the renormalized finite entropy  is quite a generic feature of the
holographic models with a gapped spectrum. We discuss in some detail the conditions that ensure this behavior, and   comment on the relation to the confinement property, making contact with the
results of \cite{kutkleb}. 

Finally, we focus on a particular  model of especial interest, namely  the case of conformal field theories (CFT) in $2+1$ dimensions with a
magnetic field and/or electric charge condensates,  holographically represented by  zero-temperature extremal
black holes with finite horizon area in Anti-de Sitter space (AdS). The relevance of this system stems from the fact that 
the extensive law, characteristic of high-temperature states, does persist all the way down to zero temperature. 

\section{Review of the  holographic  ansatz}

\noindent

The starting point of the holographic proposal of \cite{ryutaka} is the {\it replica trick} calculation of the
entanglement entropy \cite{CalabreseCardy}.  Consider a quantum field theory defined on a Hamiltonian spacetime of the form $\R \times {\rm
  X}_d$, with ${\rm X}_d$ a $d$-dimensional spatial manifold.  Given a general state $\rho$ defined in terms of the
Hamiltonian quantization on ${\rm X}_d$, let ${\rm A}$ denote a $d$-dimensional region of  ${\rm X}_d$, with smooth boundary $\pt {\rm A}$,  and consider
the inclusive density matrix obtained by tracing out degrees of freedom inside ${\rm A}$; $\rho_{\rm A} = \Tr_{\rm A}
\,\rho$. The associated entanglement entropy can be computed as \cite{CalabreseCardy} \beq\label{defrept} S_{\rm A} = -\Tr\,\rho_{\rm A}
\,\log \,\rho_{\rm A} = -{d \over dn}\Bigg|_{n=1} \, \Tr \,(\rho_{\rm A})^n \;.  \eeq The path-integral representation
of $\Tr (\rho_{\rm A})^n$ can be manipulated into a frustrated (Euclidean) partition function on a theory with $n$
copies of the original QFT, \beq\label{twistop} \Tr\,(\rho_{\rm A})^n = \left\bra \, {\bf T}_{\pt {\rm A}}^{(n)}
  \,\right\ket_{{\rm QFT}^{\otimes n}} \;, \eeq where the twist operator ${\bf T}_{\pt {\rm A}}$ is defined by introducing  the following boundary
conditions in the path integral, 
\beq\label{bcon}
\Phi_i^{(+)} = \sum_{j=1}^n \Gamma_{ij} \,\Phi_j^{(-)} \;,
\eeq
 with $\Gamma_{ij} = \delta_{i, j+1} + \delta_{n,1}$
identifying fields of consecutive copies across the set ${\rm A} $. The superscripts $(\pm)$ in (\ref{bcon}) denote the `two sides' of  ${\rm A}$, when embedded in the complete Euclidean manifold  $\mathds{R} \times X_d$. Alternatively,  for any point $P \in {\rm A}$,  $\Phi^{(+)} (P)$ is obtained from $\Phi^{(-)} (P)$ by transporting the field from the `lower side' to the `upper side' of ${\rm A}$ along a closed path of linking number one with the boundary $\pt {\rm A}$. In particular, the construction shows that
the twist operator in (\ref{twistop}) is `instanton-like', i.e. it is always   sharply
localized in the time direction.

The twist operator ${\bf T}$ is locally supported on $\pt {\rm A}$. The simplest particular case occurs in $1+1$
dimensions, where the non-local twist operator is given by a {\it bilocal} product of two standard (local)  twist operators,
i.e. (\ref{twistop}) is a two-point function. In $2+1$ dimensions, ${\bf T}$ is supported on a one-dimensional curve, just like standard Wilson and 't Hooft loop operators. Taking inspiration from the case of Wilson loops, one's natural guess for the ultraviolet contribution to  
(\ref{twistop}) is \beq\label{uvasim} \left\bra \,{\bf T}_{\pt{\rm A}}^{(n)} \,\right\ket_{\rm UV}
\sim \exp\left(-\alpha_n \,\varepsilon^{\,1-d} \;|\pt {\rm A}|\right) \;, \eeq with $\varepsilon$ a
short-distance cutoff and $|\pt {\rm A}| $ the volume of $\pt {\rm A}$ in the metric of the spatial manifold ${\rm X}_d$. Although all previous expressions hold in an arbitrary QFT in a formal sense, Eq. (\ref{uvasim})
assumes that $\varepsilon^{-1}$ is taken beyond any mass scale in the theory, so that we are close to some ultraviolet
(UV) fixed point. Since (\ref{twistop}) is almost an $n$-fold product of partition functions, we expect $\alpha_n /n $
to be finite in the $n\rightarrow \infty$ limit. Furthermore, we know that $\alpha_1 =0$, since the partition function
is not `frustrated' for $n=1$. The resulting entanglement entropy has the form \beq\label{geoment} S_{\rm A} ={d\alpha_n
  \over dn}\Bigg|_{n=1} \; {|\pt {\rm A}|\over \varepsilon^{\,d-1}} + \dots\;, \eeq where the dots stand for
less divergent terms, which will depend in general on the size and shape of ${\rm A}$ and any intrinsic mass scales of
the QFT. In Eq. (\ref{geoment}) we see the expected UV scaling of the geometric entropy, i.e. non-extensive behavior
and `area law' with respect to the entangled region \cite{enten}.

\subsection{AdS and area law}

\noindent

By direct analogy with the treatment of Wilson and 't Hooft loops in gauge theories, the authors of \cite{ryutaka}
propose to compute (\ref{twistop}) in an AdS/CFT prescription involving a minimal hypersurface with boundary data given
by $\pt {\rm A}$. This procedure can be strictly justified for the particular case of two-dimensional conformal models,
for then (\ref{twistop}) becomes a two-point function of {\it local} conformal operators. More generally, we have the ansatz
\beq\label{adsansatz} \left\bra \,{\bf T}^{(n)}_{\pt {\rm A}}\,\right\ket_{{\rm QFT}^{\otimes n}} \approx \exp\left(-c_n
  \,{\rm Vol}\,( {\bar {\rm A}})\right)\;, \eeq where ${\overline {\rm A}}$ is the minimal $d$-dimensional
hypersurface dropped inside the bulk of the AdS space, with boundary conditions $\pt {\overline {\rm A}} = \pt {\rm A}$
at the UV boundary of AdS. ${\rm Vol} \left({\bar {\rm A}}\right)$ stands for the volume of the hypersurface as induced from the
ambient AdS metric.  Computing the entropy via the replica-trick formula we find \beq\label{adsansatzen} S_{\rm A}
\approx {dc_n \over dn} \Bigg|_{n=1} \; {\rm Vol}\,\left( {\bar {\rm A}}\right) = {{\rm Vol}\,\left( {\bar
      {\rm A}}\right) \over 4G_{d+2}}\;, \eeq where the precise coefficient is fixed in \cite{ryutaka} by comparison
with known standard results in two dimensions.  Considering bulk spaces with AdS asymptotics near the boundary:
\beq\label{adsim} ds^2 /R^2 \longrightarrow u^2 \; (d\tau^2 + d{\vec x}^{\,2}) + du^2/u^2\;, \eeq any minimal
hypersurface ${\overline{\rm A}}$ with a boundary component at infinity is asymptotically perpendicular to the boundary,
with a volume divergence at large $u$ of the form \beq\label{uvads} {R^d u_\varepsilon^{d-1} \,|\pt {\rm A}|
  \over 4G_{d+2}} \propto N_{\rm eff} {|\pt {\rm A}| \over \varepsilon^{\,d-1}}\;, \eeq where $u_\varepsilon =
\varepsilon^{-1}$ and $N_{\rm eff} = R^d /G_{d+2}$ is the effective number of degrees of freedom of the conformal UV
fixed point.\footnote{For example, for models governed by a large-$N$ gauge theory we have $N_{\rm eff} \sim
  N^2$. Other models, such as the theory on a stack of $N$ M2-branes, have $N_{\rm eff} \sim N^{3/2}$, whereas $N_{\rm
    eff} \sim N^3$ for a stack of $N$ M5-branes.} Hence, the holographic ansatz obtains the expected UV structure
(\ref{uvasim}).

In addition to the UV asymptotics, one can determine a UV-finite part that is always present in the conformal
approximation of (\ref{adsim}), the renormalized entanglement entropy. To find this term, let us focus on a simple
situation where ${\rm X}_d = {\R}^d$ and let ${\rm A}$ be the $d$-dimensional {\it strip}, i.e.  the product of a finite
interval of length $\ell$ times an infinite hyperplane of codimension one in space: ${\rm A} = [-\ell/2, \ell/2] \times
{\R}^{d-1}$. By translational invariance on ${\R}^{d-1}$ we know that the entropy will be extensive along these $d-1$
dimensions. Furthermore, a rescaling of the boundary coordinates, ${\vec x}$, induces a rescaling of $\ell$ which can be
absorbed in a rescaling of $u$ in (\ref{adsim}) leaving the metric invariant. Hence, the induced volume of
${\overline{\rm A}}$ is invariant under these combined scalings and the finite part of the entropy must be of the form
\beq\label{finen} C_{\rm A} (\ell) =\ell {d \over d\ell} S_{\rm A} \propto N_{\rm eff} \,{ |{\R}^{d-1}| \over
  \ell^{\,d-1}}\;, \eeq where the factor of $N_{\rm eff}$ is ensured by dimensional analysis and the overall factor of
$G_{d+2}^{-1}$. The quantity defined in (\ref{finen}), designed to remove the leading UV term, is also called the
`entropic C-function' \cite{CasiniHuerta}. Therefore, a form of `area law' also holds at conformal fixed points, at the
level of the renormalized entropy.

If the UV fixed point is associated to a holographic model of the form ${\rm AdS}_{d+2} \times { K}$, with ${K}$ some
compact Einstein manifold of dimension $d_K = D-d-2$, this extra structure only enters (\ref{adsansatzen}) through the
Kaluza--Klein reduction of Newton's constant $G_{d+2}^{-1} = G_D^{-1} \;{\rm Vol}\,(K)$. Therefore, we can generalize
the prescription to the higher-dimensional description via a minimal $(D-2)$-dimensional hypersurface which wraps
completely the internal manifold. The Kaluza--Klein ansatz also applies to warped products of AdS-like spaces and
compact manifolds, where the radius of curvature of both factors has a non-trivial dependence on the holographic
coordinate $u$. If $R$ and $G_{d+2}$ denote the asymptotic values of curvature radius and Newton's constant, the ansatz
(\ref{adsansatzen}) remains valid, ensuring UV asymptotics controlled by $N_{\rm eff} \sim R^d /G_{d+2}$ as before, but
with modified behavior at the level of the renormalized entropy. Models arising from ten-dimensional string backgrounds
are treated in the same way, provided we remember to use the Einstein-frame metric in the ten-dimensional set up before
the Kaluza--Klein reduction.\footnote{The difference between both frames amounts to a factor of $\exp(-2\phi)$ in the
  induced volume 8-form.}

In order to fix the notation, let us consider $(d+2)$-dimensional spaces with Einstein-frame metric \beq\label{typemet}
ds^2 = R^2 \,\alpha(u)^2 \,\left( \frac{\beta(u)^2}{h(u)} \,du^2 + d{\vec x}^{\,2} + h(u)\,d\tau^2\,\right)\;, \eeq which are assumed
to asymptote AdS$_{d+2}$ at $u\rightarrow \infty$ with radius $R$, i.e. the background profile functions, $\alpha(u)/u$,
$\beta(u)\,u^2$ and $h(u)$ all approach unity at infinity. More general situations, where the model has no smooth UV
fixed point, can be dealt with by defining a running number of degrees of freedom $N_{\rm eff} (u)$. We include the
Lorentz-violating profile function $h(u)$ in order to extend the analysis to black-hole spacetimes.
 
One can further reduce the expression (\ref{adsansatzen}) for the particular situation of the strip: ${\rm A} (\ell) =
[-\ell/2, \ell/2 ] \times {\R}^{d-1}$, leading to a bulk hypersurface ${\overline {\rm A}}$ called the `straight belt'
in \cite{ryutaka}. The `longitudinal' entropy density along the strip, $s_{\rm A} = S_{\rm A} /|\R^{d-1}|$, is
given by \beq\label{eq:generic_functional} s_{\rm A} = \frac{R^d}{4 G_{d+2}} \int_{-\frac{\ell}{2}}^{\frac{\ell}{2}} dx
\, \gamma(u)\,\sqrt {1 + {(\beta(u) \partial_x u)^2 \over h(u)} } \; ,
\end{equation}
with $\gamma(u) \equiv \alpha(u)^d$. This functional defines a variational problem for the profile $u(x)$, which is
parametrized by the radial position $u_*$ of the turning point $\partial_x u=0$. The functional relation between $\ell$
and $u_*$ is given by\begin{equation}\label{equil} \ell(u_*) = 2 \gamma(u_*) \int_{u_*}^{\infty} \frac{du\,
     \beta(u)}{\sqrt{h(u)\left( \gamma(u)^2- \gamma (u_*)^2\right)}} 
\,.
\eeq
  Notice, however, that the minimizing
hypersurface might not be a smooth one, in which case (\ref{equil}) ceases to apply.

The law $\ell(u_*) \sim 1/u_*$, characteristic of the conformal case, can be readily obtained by a heuristic
argument. Let us model the surface ${\overline {\rm A}}$ as the union of two components. The first component is a
cylinder extending from radial coordinate $u_m$ up to $u_\varepsilon$, and subtending a strip of length $\ell$ in the
boundary metric.  The second component is the `cap' at $u=u_m$. The entropy functional is obtained by adding the volumes
of each component. Factoring out the longitudinal volume of the strip we find
$$
s_A (u_m) \sim N_{\rm eff} \,(u_\varepsilon^{\,d-1} - u_m^{\,d-1} \,) + N_{\rm eff}\, u_m^{\,d} \,\ell\;,
$$
with the first term coming from the cylinder and the second one from the cap at $u=u_m$. Minimizing with respect to
$u_m$ we find an optimal turning point satisfying $ (d-1)\,u_*^{\,d-2} -\ell\,d\,u_*^{\,d-1}= 0$, which yields the
desired relation $\ell u_* \sim 1$.

\section{Extensivity and infrared walls}

\noindent

In a model that combines a UV fixed point and an intrinsic energy scale $M$, the holographic dual geometry will remain
approximately AdS until we reach radial coordinates of order $u_0 \sim M$, and the form (\ref{finen}) is a good
approximation as long as $\ell M \ll 1$.  A typical instance of nontrivial infrared (IR) behavior is that the radial
coordinate `terminates' at $u_0 = M$, either because of the existence of some sort of sharp wall, a `repulsive'
singularity, or perhaps a vanishing cycle along the rest of the dimensions.
In this section we study the different types of qualitative behavior to be expected when the turning point $u_*$ approaches the wall position $u_0$. By the conformal UV/IR relation, $\ell u_* \sim 1$,  
this will occur roughly around $\ell \sim 1/M$.

\subsection{IR wall avatars}

\noindent

 The precise functional dependence $\ell(u_*)$ is specified in
Eq. (\ref{equil}), which we may rewrite as
\beq\label{equilrv}
\ell =  \int_1^\infty {dx\, \zeta(x) \over \sqrt{x^2 -1}}\;,
\eeq
where  we have made a change of variables $x= \gamma(u) / \gamma(u_*)$  and we have defined the  
function $\zeta(x)$  by 
\beq\label{zetaf}
\zeta(x) = 2{\gamma(1)\, \beta(x) \over \gamma' (x) \sqrt{h(x)}} \;,
\eeq
where $\gamma'  \equiv d\gamma /du$. The turning point $u=u_*$ lies at $x=1$ in the new variables, so that $\gamma(1) = \gamma(u_*)$, and the wall sits at some $x=x_0 < 1$. In the following, we use the notation  $\gamma(u_0) \equiv \gamma_0$ and
$\gamma' (u_0) \equiv  \gamma'_0$. There are three basic types of qualitative behavior as the turning point of {\it smooth} surfaces approaches the wall, $x_0 \rightarrow 1$, depending on whether $\ell$ diverges, approaches a constant value, or vanishes in this limit. These three alternatives can be translated into corresponding qualitative properties of the background, through the structure of the function $\zeta (x)$ in the vicinity of the wall $x\sim x_0$. 

\begin{figure}
  \begin{center}
    \epsfig{file=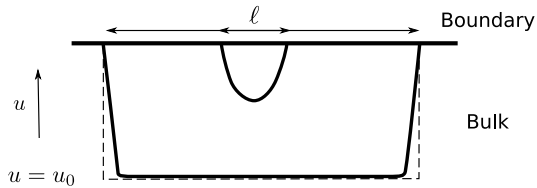, width= 7cm}
    \caption{\small Schematic plot of the typical configuration of a  smooth minimal surface that gives us the entanglement entropy
      in the presence of infrared walls. Close to the position of the infrared wall, the surface prefers to lean on
      it, giving an extensive contribution to the entropy. We also show the `capped cylinder'  in dashed lines, which becomes a better and better approximation to the minimal surface as $\ell \rightarrow \infty$.}
  \end{center}
\end{figure}

In our analysis, we shall adopt some technical assumptions derived from experience with concrete examples of holographic duals. First, we assume that $\beta(u)$ is smooth and positive for all  $u\geq u_0$.  The Schwarzschild-like factor $h(u)$ will be positive as well, except for a possible single zero at $u=u_0$. Finally,  the warp factor $\gamma(u)$ is  taken to be positive and monotonically increasing for all $u>u_0$, but we allow for the possibility that it may vanish right at the wall position. Within these technical assumptions, the only possible singularity structure of $\zeta(x)$ is either a pole at $x=x_0$, coming from a zero of the warp-factor derivative, $\gamma' (u_0)=0$, or a square root singularity $\zeta(x) \sim (x-x_0)^{-1/2}$, coming from a regular black-hole horizon, 
$h(u_0) =0$. 

Since we are assuming that the model has an UV fixed point, we know that the contribution of large values of $x$ to (\ref{equilrv}) is well approximated by the conformal law $\ell_{\rm high} 
\sim 1/u_0 \sim 1/M$.  Then, if $\zeta(x)$ is smooth in the vicinity of $x=1$ we have a finite $\ell \propto 1/u_0$ as $u_* \rightarrow u_0$,
perhaps decorated with some numerical factors involving dimensionless parameters of the background. A vanishing
$\ell$ in the wall limit can only result from a vanishing warp factor $\gamma_0 =0$.

Finally,  when $\zeta(x)$ diverges as $(x-x_0)^{-1/2}$ or faster, we have 
$\ell \rightarrow \infty$ as  $x_0 \rightarrow 1$. Then we say that the smooth surface {\it saturates} at the wall. In this case we have the situation depicted in Fig. 1, where a smooth surface asymptotically rests over the wall, forming a  cap that is smoothly joined by an almost cylindrical surface of section $\pt {\rm A}$, reaching out to the boundary. 
In the limit of very large $\ell$ the resulting entropy is very well approximated by that of the `capped cylinder', obtained by adjoining a `cap' $u_0 \times {\rm A}$ to the cylinder $[u_0 , \infty] \times \pt {\rm A}$ (in the case of the strip, the cylinder is just given by the two straight $[u_0 , \infty] \times \mathds{R}^{d-1}$ panels, 
suspended a distance $\ell$ apart).

We are now ready to detail the casuistics of different types of walls regarding their saturation properties.

\subsubsection{Soft walls}

\noindent

\begin{figure}
  \begin{center}
    \epsfig{file=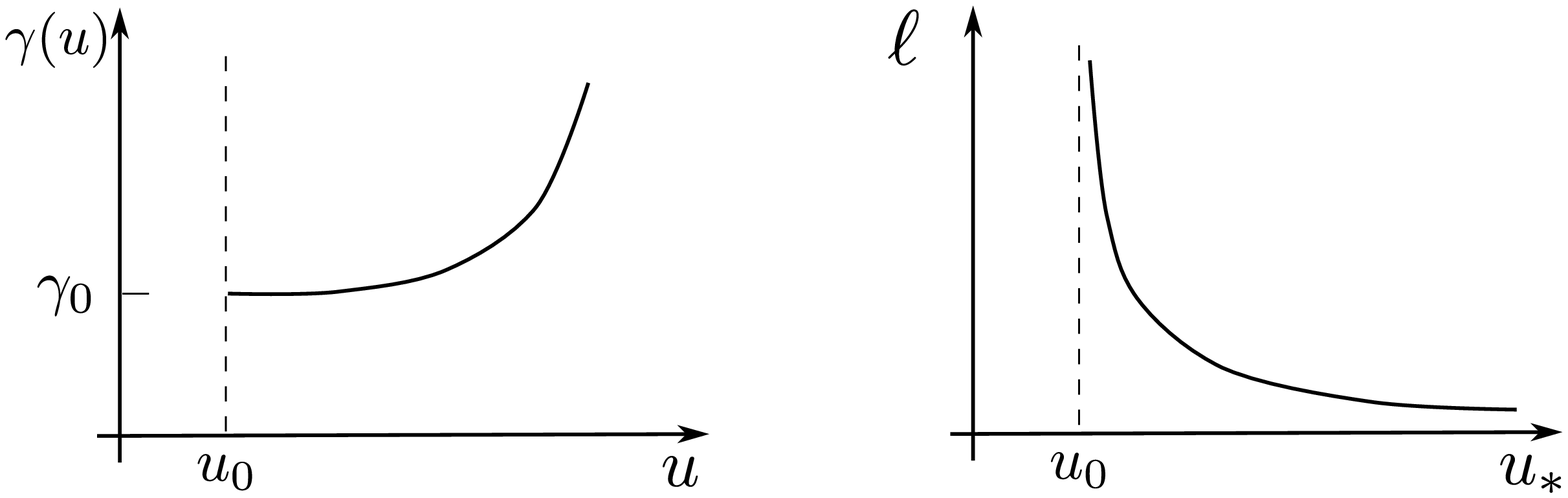, width= 10cm}
    \caption{\small Qualitative behavior of the warp factor near a soft wall, with $\gamma'_0 =0$ and $\gamma_0 >0$. On the right,  the corresponding UV/IR relation with  saturation, $\ell\rightarrow \infty$, at the wall.}
  \end{center}
\end{figure}

These walls are characterized by a local minimum of the warp factor, $\gamma'_0 =0$, with $\gamma_0 >0$ (we set $h(u) =1$ for simplicity). Then $\zeta(x) \sim (x-x_0)^{-1}$ and the smooth surfaces saturate (cf. Fig 2), $\ell \rightarrow \infty$ as $u_*\rightarrow u_0$. We call these walls `soft' because the function $\gamma(x)$ usually extends for $x<x_0$, albeit with negative derivative. Static masses feel a repulsive gravitational force that prevents physical probes from reaching far into the $x<x_0$ region. Particular examples include the confinement models of \cite{dilcon} based on backgrounds with naked singularities. Hence, these models are usually regarded as formal qualitative tools at best.

\subsubsection{Hard walls}

\noindent

These walls are defined by $\gamma_0, \gamma'_0 >0$, i.e. they behave formally as the `sharp cutoff' regularization of a simple AdS background and are commonly used in phenomenological discussions of AdS/QCD, as in \cite{strss}. In a formal sense, they can be obtained as the `stiff' limit of the soft walls in the previous subsection, that is to say the limit in which the minimum of $\gamma(u)$ at $u=u_0$ is made sharper and sharper. 

Regarding the entanglement entropy, the most striking property of these walls  is a UV/IR relation $\ell(u_*)$ with a maximal value of $\ell$ (cf. Fig. 3). Indeed,  for $\ell > \ell_{\rm max}$ the minimizing surface is not smooth.  By explicit inspection, we can see that the cylinder $[u_0 , \infty] \times \pt {\rm A}$, with a cap $u_0 \times {\rm A}$ right at the wall
is a minimal surface (any perturbation takes an element of the surface to larger $u$, so that it increases the induced volume). Notice that this surface is not smooth because it has sharp edges at the juncture between the cap and the cylinder, but it is clearly connected and has the same topology as the smooth surfaces that extremize the problem
for $\ell < \ell_{\rm max}$.

\subsubsection{Resolved walls}

\noindent

These walls are characterized by a vanishing warp factor, $\gamma_0 =0$,  at the location of the wall. As a result,
the function $\ell(u_*)$ vanishes as $u_* \rightarrow u_0$. As depicted in Fig. 4 there is again a maximal $\ell_{\rm max}$ but now there are two smooth extremal surfaces for any $\ell<\ell_{\rm max}$. This is the case studied at length in the examples of Ref. \cite{kutkleb}.  It corresponds to most well defined models of confinement, in which the vanishing induced metric $\gamma_0 =0$ arises from a vanishing cycle in some extra dimension, in such a way that the higher-dimensional metric is completely smooth at $u=u_0$. For this reason we refer to these walls as `resolved' by the higher-dimensional uplifting. 

\begin{figure}
  \begin{center}
    \epsfig{file=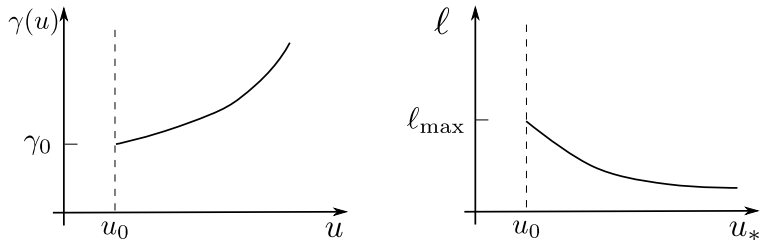, width= 10cm}
    \caption{\small Qualitative behavior of the warp factor near a hard wall, with $\gamma'_0 >0$ and $\gamma_0 >0$. On the right,  the corresponding UV/IR relation with a maximal value of $\ell$.}
  \end{center}
\end{figure}

In addition to the mentioned examples studied in \cite{kutkleb}, there is a different type of model that belongs to this class, namely that of spherical distributions of D-branes in the Coulomb branch of $N=4$ super Yang--Mills theory (cf. \cite{trivedi}).  The dual supergravity configuration consists of a standard ${\rm AdS}_5 \times {\rm S}^5$ background for
$u>M$, matched  with  a flat ten-dimensional metric inside the sphere at $u=M$, where $M$ is the mass scale set by the Higgs mechanism in the Coulomb branch. The hypersurface entering the
entropy ansatz wraps the ${\rm S}^5$ spheres, which remain of constant volume for all $u>M$ but shrink to zero size
as a standard angular sphere in polar coordinates for $u< M$. Since the five-dimensional warp factor $\gamma(u)$ is proportional to the volume of the internal manifold at fixed $u$, we find ourselves in the qualitative situation of Fig. 4.

Among the two solutions for each $\ell < \ell_{\rm max}$, the  standard minimal embeddings correspond to the solution with the larger value of $u$, whereas the branch of solutions with $\ell \rightarrow 0$ as $u\rightarrow u_0$ corresponds to locally unstable surfaces. The absence of stable minimal surfaces for $\ell > \ell_{\rm max}$ makes this case similar to that of the hard wall, and the problem is resolved in the same fashion. Namely, the minimal surface at very large $\ell$
is not smooth, consisting of the capped cylinder lying right at the wall. \footnote{As shown in \cite{kutkleb} the capped cylinder remains the absolute minimum down to $\ell_c$, a critical length of order $\ell_{\rm max}$, albeit somewhat smaller.} A crucial difference with respect to the hard
wall case is that $\gamma_0 =0$ and now the induced volume of the cap is not just minimal, but in fact it
vanishes. Hence, the contribution of the minimal surface to the entropy is just given by the cylindrical portion, which was referred to in \cite{kutkleb} as the `disconnected surface', in the particular case of the strip. 

It is important to stress, however, that the minimal surface with degenerate cap is just a member of a continuous set of surfaces with standard topology and with induced volume arbitrarily close to the minimal one. For this reason we do not
think that the picture of a `disconnected surface' is relevant or appropriate in general, despite the fact that it gives the same answer as the capped cylinder for the particular case of these walls (more on this in a subsequent section).

\subsubsection{Thermal walls}

\noindent

Finally, the remaining case with a clear physical interpretation is that of a black hole horizon, 
$h(u_0) =0$, corresponding to a plasma phase in the dual theory on the boundary. A regular (non-extremal) horizon will have $h'(u_0) \neq 0$ and the effective function diverges as
$\zeta(x) \sim (x-x_0)^{-1/2}$ in the vicinity of the wall. This translates into a logarithmically divergent  $\ell$ in the limit $u_* \rightarrow u_0$, as defined by (\ref{equilrv}). In other words, we have saturation by smooth surfaces and a behavior qualitatively similar to that of the soft walls. 

The full Euclidean metric (\ref{typemet}) of the  black hole backgrounds is  smooth at the wall, since the
compact thermal cycle with the inverse temperature identification $\tau \equiv \tau + \beta$,  attains vanishing size
at $u=u_0$ in a smooth way. In this sense, this metric is similar to the higher-dimensional versions of the metric in
the case of the resolved walls, discussed in the previous subsection. It is very important, however,  to distinguish the two situations by recalling that the embedded surface ${\overline{\rm A}}$ is {\it always} localized in the $\tau$ direction, so that the vanishing thermal cycle does not translate in a vanishing induced metric at the wall, i.e. we still have $\gamma_0 >0$ in this case. Conversely, the vanishing cycles in the examples of the previous subsection are being wrapped by the minimal surface, and do contribute to the vanishing of $\gamma_0$ there.

\begin{figure}
  \begin{center}
    \epsfig{file=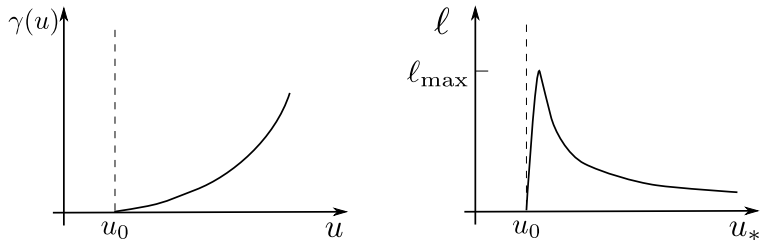, width= 10cm}
    \caption{\small Qualitative behavior of the warp factor near a resolved wall, with $\gamma'_0 >0$ and $\gamma_0 >0$. On the right,  the corresponding UV/IR relation with a maximal value of $\ell$ and two smooth extremal surfaces for each $\ell < \ell_{\rm max}$.}
  \end{center}
\end{figure}

\subsection{Extensivity of the entropy}

\noindent

We conclude that in all cases the minimal surfaces effectively saturate at the wall. In some cases the saturating surface
is smooth and well described by Eq. (\ref{equilrv}), whereas in some others the minimal surface is not smooth, given by
the `capped cylinder' at the wall. In either case, the entropy is well approximated by that of the capped cylinder, and in the cases  of non-smooth minimal surfaces, it is exactly given (in the classical limit) by the volume of the capped cylinder. In fact, we may carry the complete discussion at a qualitative level in terms of a restricted family of surfaces, corresponding to  cylinders of base $\pt {\rm A}$  with a cap  at some  $u=u_m \geq u_0$.  The resulting entropy takes the form
\beq\label{cili}
S(u_m) = {1\over 4} N_{\rm eff} \left[ |{\rm A} | \, \alpha(u_m)^d  + |\pt {\rm A} |  \int_{u_m}^{u_\varepsilon} {du \over \sqrt{h(u)}} \beta(u) \alpha(u)^{d} \right]\;,
\eeq
where the first term is the contribution of the cap at $u=u_m$ and the second term corresponds to the cylinder extending from $u=u_m$ up to the cutoff scale $u=u_\varepsilon$. Recall that $|{\rm A}|$ and
$|\pt {\rm A}|$ denote  the volumes of ${\rm A}$ and its boundary  in the field-theory metric. For the case of the strip $2 |{\rm A}| = \ell |\pt {\rm A}|$.  Taking the first derivative we find 
\beq\label{cilid}
{dS \over du_m} = {N_{\rm eff} \over 8} |\pt {\rm A}| \left[ \ell \, d\,\gamma' (u_m) - 2{\beta (u_m) \gamma(u_m)^{} \over \sqrt{h(u_m)}}\right]\;,
\eeq
where we have made use of $\gamma(u) \equiv \alpha(u)^d$. Furthermore, equating it to zero fixes $u_m= u_*$ and we obtain  the
UV/IR relation in this approximation, 
\beq\label{cuvi}
\ell(u_*) = {2 \over d} {\beta (u_*) \over \sqrt{h(u_*)}} {\gamma(u_*)^{} \over \gamma' (u_*)}\;,
\eeq
 Eq. (\ref{cuvi}) shows all the qualitative
features discussed previously on the basis of (\ref{equilrv}). In particular, we see that saturation, i.e.
$\ell(u_* \rightarrow u_0) \rightarrow \infty$ occurs whenever $h(u_0) =0$ or $\gamma' (u_0) =0$, whereas $\gamma(u_0)=0$ ensures that there is a branch of solutions with $\ell \rightarrow 0$ in the wall limit.  Eq. (\ref{cilid}) also shows that  for resolved walls and hard walls with $\gamma'_0 >0$, the entropy functional  at fixed and large $\ell$ satisfies $dS/du_m >0 $ at $u_m =u_0$, so that the entropy is locally minimized by the capped cylinder at $u_m = u_* =u_0$ provided $\ell$ is large enough.

Hence, the entropy at very large $\ell$ is well approximated by the induced volume of the capped-cylinder in all cases. The contribution from the cylinder  contains the cutoff dependence and it is proportional to $|\pt {\rm A}|$, i.e. it gives an area law. On the other hand, at large $\ell$ it is asymptotically independent of $\ell$, so that it drops 
 from the calculation of
$C_{\rm A} (\ell)$. The remaining finite term is the volume of the cap at $u_* \approx u_0 =M$, appropriately
redshifted to that radial position, i.e.  
\beq\label{capen}
S_{\rm cap} \approx {R^d \,\gamma_0\, |{\rm A}| \over 4G_{d+2}} \;.
\eeq
 For a
model asymptotic to AdS with curvature radius $R$, we can multiply and divide by $(u_0)^d$ to obtain 
the extensive law
\beq\label{finalex}
C_{\rm A} (\ell) \approx S_{\rm cap} \propto \eta_M\; N_{\rm eff} \; M^{d} \;|{\rm A}|\;, \eeq
 with $N_{\rm eff }$ the effective
number of degrees of freedom at the UV fixed point CFT and $M = u_0$, the value of the capping coordinate, determining
the scale of the mass gap. The new parameter \beq\label{newpa} \eta_M = {\gamma_0 \over (u_0)^d} \eeq keeps track of the
`flow' of relevant parameters from the UV fixed point at $u=\infty$ down to the IR mass scale $u_0 = M$.  For
example, for a model obtained by Kaluza--Klein reduction on some internal manifold $K$, we have \beq\label{partcaseeta}
\eta_M = {{\rm Vol} (K_0) \over {\rm Vol} (K_\infty) }\;, \eeq where all volumes must be computed in Einstein-frame
conventions, in the case of string backgrounds. In fact, we could define an effective `infrared' number of degrees of
freedom by the product $N_{{\rm eff},0} = \eta_M \,N_{\rm eff}$.

In many situations the flow parameter $\eta_M$ is not qualitatively important. This is for example the
case of black-hole backgrounds, with a dual interpretation in terms of thermal states of the QFT. In general $\gamma_0 \neq 0$ at such
a black hole horizon, and we obtain an extensive component of the entropy, in accordance with general expectations
(c.f. \cite{ryutaka}) \beq\label{ftem} S_{\rm A} (T) \propto N_{\rm eff} \,T^{\,d} \,|{\rm A}|\;, \eeq where
$u_0 \sim T$ for a large-temperature black hole in AdS. In the next section we introduce a very interesting system
in which magnetic fields or charge condensates are capable of supporting a zero-temperature horizon with
the formal properties of a thermal wall, namely supporting an nontrivial extensive law for the entanglement entropy.

Finally, the most notable exception to the extensive behavior is the case of resolved walls \cite{kutkleb}, with $\gamma_0 =0$, which automatically yield $\eta_M =0$ by the vanishing of ${\rm Vol} (K_0)$ in (\ref{partcaseeta}).

\subsubsection{The role of disconnected surfaces}

\noindent

We have stressed that the `disconnected surfaces' invoked in \cite{kutkleb} are in fact standard connected surfaces in which the `endcap' contributes zero volume due to the degenerate induced metric at the wall.\footnote{The term `disconnected surface' strictly refers to the case of the strip, in which $\pt {\rm A}$ is the union of two $\mathds{R}^{d-1}$ planes. More generally, the `disconnected surface' is regarded as   the
cylinder $[u_0 , \infty] \times \pt {\rm A}$.} Indeed, there is a continuous family of surfaces of standard topology that approximate arbitrarily well the volume of the so-called disconnected surface.  Exactly the same situation occurs in the two-point function of Polyakov loops, using in that case  the vanishing circle Euclidean black hole backgrounds (cf. for example the discussion in \cite{oogurigross}).\footnote{We stress once more that the surfaces computing entanglement entropy are always orthogonal to thermal circles, somewhat like {\it spatial} Wilson lines and strictly {\it unlike} timelike Wilson lines, also known as Polyakov lines.} 

 Invoking disconnected surfaces
has the additional problem that they would naturally minimize the large $\ell$  entropy in the presence of   
 thermal walls, a situation in which we would lose the understanding of the extensivity of thermal entanglement
 entropy. We believe it more likely that disconnected surfaces are related to  a different observable, namely the
 so-called {\it mutual information} (cf. for example \cite{chuang} for a discussion of its properties and uses in quantum
 information). Mutual information between a bipartite partition of a system is defined as
 \beq\label{mutin}
 I[{\rm A},{\rm B}] \equiv S_{\rm A} + S_{\rm B} - S_{{\rm A} \cup {\rm B}}\;,
 \eeq
 where the last term is the total von Neumann entropy of the whole system, and it vanishes for a pure state. On the other hand, for a thermal state the extensive terms cancel out of (\ref{mutin}) and one is left with the area-law terms sensitive to the UV cutoff. Applying this prescription to the holographic ansatz in a black hole background one finds that the `caps' in the saturation surfaces are subtracted out by the
 standard Bekenstein--Hawking entropy  and what remains is twice the volume of the uncapped cylinder, scaling with an
 area law. This prediction is in agreement with the work of Ref.~\cite{ciracm} where they demonstrate for lattice systems that the mutual
 information follows an area law. In this respect we propose to reinterpret the work in \cite{pando-zayas} as a
 calculation of the mutual information rather than the strict entanglement entropy at finite temperature.

\subsubsection{Phase transitions}

\noindent

In the previous subsections we have discussed the different qualitative possibilities corresponding to different types of walls.  With little more effort one can extend the analysis further and consider combined situations, where for instance a soft wall, with $\gamma'_0 =0$, develops a degenerate metric at the wall position, $\gamma_0 \rightarrow 0$. In this case,
one must study the relative rates at work, and concludes that the behavior of the prefactor $\gamma(1)$ dominates, producing a situation akin to that of resolved walls. 

Another interesting complication is the consideration of systems with IR walls {\it and} finite temperature. In this case one must compare the competing effects of  the warp factor $\gamma(u)$ {\it versus} the Schwarzschild factor $h(u)$. One must also consider phase transitions of the background, or Hawking--Page type. Roughly speaking the dominating background at a given temperature is the one with the largest value of the wall coordinate (up to transient effects that depend on the precise value of the free energies).  For $T\ll M$ the system is characterized by a zero-temperature wall and $h(u)=1$ (the `confined phase'), whereas for $T\gg M$ the system is always in the plasma phase, and the relevant wall is the thermal one, determined by the zero of $h(u_0) =0$. 

We may then consider the behavior of the function $s_{\rm A} (\ell, T, M)$. For $\ell^{-1} \gg T, M$ the entanglement entropy
shows conformal behavior. On the other hand, for $\ell^{-1}  <{\rm max} (T, M)$ the minimal surface will be well-approximated by the capped cylinder, and the finite part of the entanglement entropy will scale `extensively' with the law ({\ref{ftem}) for $T\gg M$ or the law (\ref{finalex}) for $T\ll M$. In the cold phase the actual behavior will depend on whether
$\eta_M$ vanishes or not. If $\eta_M =0$, such as the case of resolved walls, the entanglement entropy at very large $\ell$ makes a ${\cal O}(N_{\rm eff})$ jump across the thermal phase transition. On the other hand, for soft or hard walls with
$\eta_M \sim 1$ the large-$\ell$ entropy is extensive at {\it both} low and high temperatures and only the numerical coefficient, proportional to $\eta_M$, may jump discontinuously across the deconfining phase transition.

\section{A cold case example}

\noindent

The characteristic example of an extensive renormalized geometric entropy is that of a highly thermalized state, which
is modeled by a large AdS black hole in the holographic map. In this picture, the condition for extensivity is the
occurrence of an effective wall at $u=u_0$ which saturates the growth of the minimal hypersurface as $\ell \rightarrow
\infty$.  In the zero-temperature limit, $u_0 \sim T \rightarrow 0$ and the extensive term vanishes. In this section we
present an example of a system that maintains the extensivity down to zero temperature, provided we have a magnetic
field and/or electric charge densities on the vacuum, supported by a large number of degrees of freedom.

As an explicit example with specific interest we can focus on the extremal dyonic black hole in AdS$_4$, studied
recently in \cite{kovtun}. This black hole has a zero-temperature horizon with finite entropy density, supported by
magnetic and electric fluxes. The holographic dual corresponds to a  maximally supersymmetric conformal field theory in $2+1$ dimensions, the worldvolume theory of a stack of $N$ M2-branes, in the presence of an
external magnetic field and an electric charge chemical potential, both of them associated to a gauged $U(1)$ subgroup
of the $SO(8)$ R-symmetry. The dyonic black hole is a solution of the Einstein--Maxwell theory on AdS$_4$ emerging as a
consistent truncation of eleven dimensional supergravity on ${\rm AdS}_4 \times {\rm S}^7$ \cite{mth}.

The geometry is given by
\begin{equation}
  ds^2  = R^2\,u^2 \left( h(u) d\tau^2 + dx_1^2 + dx_2^2 \,\right) + \frac{R^2\,du^2}{u^2 \,h(u)}  \;, 
\end{equation}
with
\begin{equation}
  h(u) = 1 + (h^2 + q^2) \frac{u_0^4}{u^4} - (1 + h^2 + q^2) \frac{u_0^3}{u^3} \;,
\end{equation}
and electromagnetic field
\begin{equation}
  F = B \,dx_1 \wedge dx_2 +\mu \,u_0\,dt\wedge du^{-1}  \;,
\end{equation}
where the dimensionless electric and magnetic parameters $h, q$ are related to the magnetic field $B$ and the charge
chemical potential $\mu$ in the dual CFT by the relations \beq\label{transl} B = h \,u_0^2\;, \qquad \mu = - q \, u_0
\;.  \eeq The Hawking temperature of this black hole is given by \beq\label{hawkt} T = {u_0 \over 4\pi} \left(3-h^2 -
  q^2 \right)\;.  \eeq The crucial property for our purposes is that these black holes admit a zero-temperature limit
at finite $u_0$, provided $h^2+q^2 \rightarrow
3$.  Note that the restriction $h^2+q^2 = 3$ is perfectly compatible with general independent values of $\mu$ and $B$. In this limit, the horizon position becomes a function of the magnetic field and the chemical potential through \beq\label{linkbq} u_0^2 \equiv
M_{\rm eff}^2 = {1\over 6} \left[ \mu^2+ \sqrt{ \mu^4 + 12 B^2 } \right]\;.  \eeq The resulting model satisfies the
conditions laid down in the previous section. There is an effective mass scale determined by the charge and magnetic
condensate, $M_{\rm eff}$, which marks the threshold separating the conformal behavior of the entropy,
\beq\label{conb}
s_{\rm A} \sim N_{\rm eff} \left( {1\over \varepsilon} - {1\over \ell} \right)
\eeq
from the extensive behavior
\beq\label{exb}
s_{\rm A} \sim N_{\rm eff} \left(  {1\over \varepsilon} +  M_{\rm eff}^2 \, \ell \right)\;.
\eeq

One can check the preceding statements by calculating numerically the entanglement entropy per unit length in the
direction parallel to the strip. More precisely we find \footnote{The absolute normalization follows from the relation 
  $G_4 = 3 R^2 /\sqrt{8N^3}$ which gives us in this case $N_{\text{eff}}=
  \sqrt{2 N^3} /3$.}
\begin{equation}\label{deff}
  s_{\rm A} = \frac{\sqrt{2}}{3} N^{\frac{3}{2}} \left\{ \frac{1}{\varepsilon} + f(\ell) \right\} \;,
\end{equation}
 where $f(\ell)$ is cutoff independent and interpolates between the two extreme limits (\ref{conb}) and (\ref{exb}), as
 shown in  Fig.~\ref{fig:dyonic_bh}. The explicit form of $f(\ell)$ is given parametrically by
 \begin{equation}
   f(u_*) = u_*\left\{ -1 +  \int_0^1 dv \left( \frac{1}{v^2\sqrt{h(v)}\sqrt{1-v^4}} - \frac{1}{v^2} \right) \right\}
   \;,\quad
   \ell(u_*) = \frac{2}{u_*} \int_0^1 \frac{v^2 dv}{\sqrt{h(v)}\sqrt{1-v^4}}
 \end{equation}
where $$h(v) = 1 + 3\frac{u_0^4}{u_*^4}v^4 - 4 \frac{u_0^3}{u_*^3}v^3 \;.$$

These results can be generalized to arbitrary CFTs in dimension $d+1$, with `cold' charge condensates dual to
extremal charged  black holes in AdS$_{d+2}$ (magnetic fields appear together with the charge only for
$d+2 =4$). The corresponding supergravity solutions are studied in Ref. \cite{roberto}.  In these cases, the large
volume asymptotics of the geometric entropy is controlled by the chemical potential: \beq\label{ultim} S_{\rm A} \sim
N_{\rm eff} \,M_{\rm eff}^{\,d} \,|{\rm A}|\;, \eeq with $M_{\rm eff} = \mu$.

\begin{figure}
  \begin{center}
    \epsfig{file=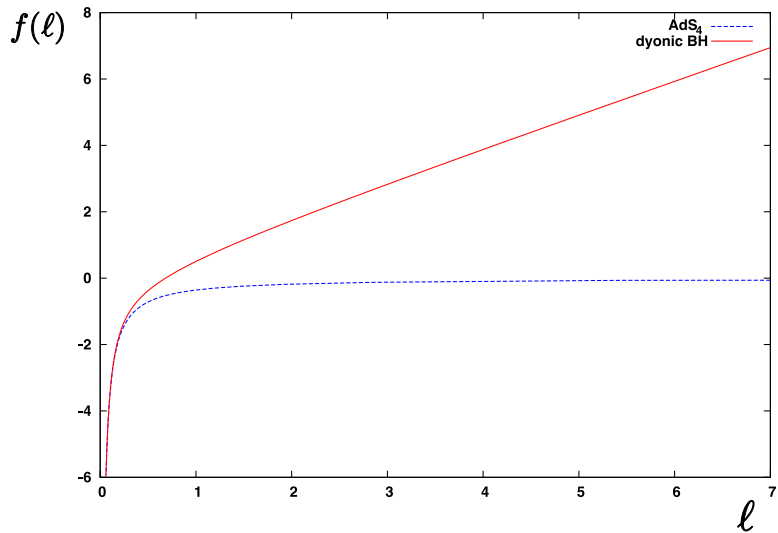, height= 7cm}
    \caption{\small Numerical plot of the function $f(\ell)$,  as defined in Eq.~(\ref{deff}), in units where $u_0 = M_{\rm eff} =1$. The red continuous line corresponds to the case of a CFT in $2+1$
      dimensions in the presence of an external magnetic field and/or charge condensate, with a bulk dual given by 
      the  dyonic
      black hole background. The blue dashed line corresponds to the same CFT without any magnetic field or charge
      condensate, as given by the AdS$_4$ dual background. We see how  extensive behavior, linear in $\ell$,  sets it at $M_{\rm eff} \ell \gg 1$ in the presence of  magnetic field and/or charge condensate.}
    \label{fig:dyonic_bh}
  \end{center}
\end{figure}

In the preceding discussion we saw how the radius of the black hole vanishes as $B$ and $\mu$ are taken to zero,
Eq. (\ref{linkbq}). Therefore, the entanglement entropy $s_{\rm A} (\ell, \mu)$ at fixed $\ell$ also interpolates
smoothly between the two asymptotic laws (\ref{conb}) and (\ref{exb}) as $\mu$ is turned on.  One may wonder whether
this smooth transition is maintained in other situations, in particular in the case where the CFT lives on a sphere
${\rm S}^{d}$ of radius $R_S$, dual to asymptotic AdS geometries in global coordinates \cite{roberto}.  The main effect
of the finite radius is to set a minimal value of chemical potential, $\mu_{c} \sim 1/R_S$, below which no extremal
black holes are found.  This transition is smooth in the sense that the black hole, present for $\mu>\mu_{c}$, still
grows from {\it zero size}, according to the formula
$$
u_0 (\mu) =M_{\rm eff}= {\sqrt{d^2 -1} \over (d+1) R_S} \,\sqrt{{\mu^2 \over \mu_{c}^2}  -1} \;,
$$
where $\mu_{c}^2 = d/(d-1)R_S^2$.  The phase transition between pure AdS and charged black holes becomes more severe as
soon as we depart from zero temperature.  Switching on the temperature parameter we generate a critical line in the $(T,
\mu)$ plane, \beq\label{critl} \left( {T \over T_{c}}\right)^2 + \left( {\mu \over \mu_{c}}\right)^2 = 1\;, \eeq
interpolating between our smooth $\mu=\mu_{c}$ critical point at $T=0$ and the standard $\mu=0$ Hawking--Page transition \cite{hpage}
at $T=T_c = d/2\pi R_S$.  All $T>0$ phase transitions on the critical line (\ref{critl}) are of first order, i.e.  as we
increase $\mu$ at fixed $T$, the black hole nucleates with a {\it finite} size $u_0 (T)= 2\pi T/d$. Hence, a
discontinuous behavior of $s_A$ as a function of $\mu$ will only occur at $T>0$, at least within the classical
approximation, to order ${\cal O}(N_{\rm eff})$.

\section{Concluding remarks}

\noindent

Our main observation in this note is the recognition of the extensivity as a generic property of renormalized
entanglement entropy at zero temperature, when calculated using the AdS/CFT ansatz of Ref. \cite{ryutaka}.  Whenever the
holographic geometry shows an infrared `wall' that effectively puts an end to the spacetime, the very large volume
asymptotics of the geometric entropy shows extensive behavior {\it in principle}. In practice, many IR walls in concrete
holographic models are associated to vanishing cycles in extra dimensions. In all these cases, as explicitly pointed out
in \cite{kutkleb}, the extensive term in the entropy vanishes to leading order in the $1/N_{\rm eff}$ expansion (this
was characterized in the text as $\eta_M =0$).  It was suggested in \cite{kutkleb} that the vanishing of the renormalized
entanglement entropy can be regarded as an order parameter for confinement, as all of these models hold that
property. Confinement, or finite string tension, arises in a rather similar fashion from the geometrical point of view,
namely the saturation of Wilson loops at an IR wall is very similar to the saturation of minimal hypersurfaces of
\cite{ryutaka}. There are subtle differences though, since Wilson loops are orthogonal to vanishing cycles in extra
dimensions, directly responsible for $\eta_M =0$.  
On the other hand, in high-temperature plasma phases the extensive behavior of the minimal hypersurfaces resembles
that of {\it spatial} Wilson loops.

In the better-defined models, namely the resolved walls and the thermal walls, one can say that 
the extensivity is associated to the breaking of conformal symmetry without the generation of a mass gap, since the plasma has ${\cal O}(N_{\rm eff})$ massless degrees of freedom while, in contrast, the
resolved walls have a mass gap and the extensive term in the entropy vanishes.

It is worth stressing that these comments only apply to leading order in the large $N_{\rm eff}$ expansion. There are
many examples of confinement models {\it without} strict mass gap, having Goldstone bosons in the spectrum from some
spontaneously broken global symmetry (cf. for example \cite{Gubser}). However, in all these cases, the number of massless
modes is $\mathcal{O}(1)$ in the large $N_{\rm eff}$ limit, and thus their effects are invisible in the classical
approximation to the geometric entropy that is being used here.

The typical example of extensive renormalized entanglement entropy is that of highly thermalized states. We have shown
in this paper that one can find systems where the extensivity holds down to zero temperature, supported by magnetic
field and/or charge condensates. This is an interesting prediction of the AdS/CFT ansatz for entanglement entropy,
although the microscopic interpretation of this property in weak coupling remains an open problem. Indeed, it would be
very interesting to check the conditions for this extensivity using purely QFT methods. Finally, the identification of
systems with peculiar entropy behavior at zero temperature opens up possible applications to the emerging field of
quantum phase transitions \cite{sachdevbook}.

\vspace{0.2cm}
{\bf Acknowledgments}

\vspace{0.2cm}

We are indebted to J. I. Cirac, C. G\'omez and S. Sachdev for discussions.  
This work was partially supported by MEC and FEDER
under grant FPA2006-05485, CAM under grant HEPHACOS P-ESP-00346 and the European Union Marie Curie RTN network under
contract MRTN-CT-2004-005104. C.A.F. enjoys a FPU fellowship from MEC under grant AP2005-0134.



\end{document}